\documentclass[pr,12pt,notitlepage,longbibliography]{revtex4-1}
\usepackage[T2A]{fontenc}
\usepackage[cp1251]{inputenc}
\usepackage[english]{babel}
\makeatletter\AtBeginDocument{\let\@elt\relax}\makeatother
\usepackage{amsmath}
\usepackage{bm}
\usepackage{amsfonts}
\usepackage{amssymb}
\usepackage{makeidx}
\usepackage[unicode=true,colorlinks=true,citecolor=blue,urlcolor=blue]{hyperref}
\usepackage[pdftex]{graphicx}
\usepackage{color}
\usepackage{dsfont}
\usepackage{braket}

\newcommand{\e}{\mathrm{e}}

\renewcommand{\i}{{\rm i}}
\renewcommand{\d}{\mathrm d}
\renewcommand{\braket}[1]{\left\langle #1 \right\rangle}

\begin{document}

\newcount\timehh  \newcount\timemm
\timehh=\time \divide\timehh by 60
\timemm=\time
\count255=\timehh\multiply\count255 by -60 \advance\timemm by \count255

\title{Nuclear spin dynamics and noise in anisotropic large box model\footnote{This is a translation of the original manuscript in Russian available in the supplementary files at arXiv
and at \href{https://www.dropbox.com/s/5157hwjohbj38mv/CSM_diag.pdf?dl=1}{this URL}.}}

\author{A. V. Shumilin}
\author{D. S. Smirnov}
\email[Electronic address: ]{smirnov@mail.ioffe.ru}

\affiliation{Ioffe Institute, St. Petersburg, Russia}

\begin{abstract}
  We consider the central spin model in the box approximation taking into account external magnetic field and anisotropy of the hyperfine interaction.
  From the exact Hamiltonian diagonalization we obtain analytical expressions for the nuclear spin dynamics in the limit of many nuclear spins.
  We predict the nuclear spin precession in zero magnetic field for the case of anisotropic interaction between electron and nuclear spins.
  We calculate and describe the nuclear spin noise spectra in the thermodynamic equilibrium.
  The obtained results can be used for the analysis of the nuclear spin induced current fluctuations in organic semiconductors.
\end{abstract}


\maketitle
\tableofcontents

\section{Introduction}

Charge carriers in semiconductors can be localized in many ways: at impurities, at defects, in quantum dots, at roughnessies of the quantum well interfaces, at organic semiconducting molecules~\cite{ivchenkopikus}. The most interesting case is the localization in quantum dots, when the parameters of localizing potential can be tuned over the wide range.

The first quantum dots (nanocrystals) were synthesized by Ekimov and Onushchenko in 1981 by colloidal synthesis~\cite{Ekimov1981_eng}. Theoretically the optical interband transitions in them were described by Efros~\cite{EKIMOV1985921}. A decade later in the pioneering work~\onlinecite{PhysRevLett.74.4043} the GaAs-based quantum dots were obtain by Stranski -- Krastanov method, which demonstrated narrow luminescence lines.

The interest in the investigation of the quantum dots in the XXIst century is related mainly to the spin effects~\cite{book_Glazov,singleSpin}. The spin dynamics of electrons in quantum dots in small magnetic fields is determined, as it was shown by Merkulov, Efros, and Rosen, by the hyperfine interaction with a large number of the host lattice nuclear spins~\cite{merkulov02}. This interaction leads to a number of bright spin phenomena such as Hanle effect~\cite{dyakonov_book}, polarization recovery~\cite{PRC}, resonance spin amplification~\cite{Kikkawa98}, mode locking~\cite{A.Greilich07212006}, spin inertia~\cite{PhysRevB.98.125306,PhysRevB.98.121304}, low frequency spin noise~\cite{Smirnov-review,Occupancy-noise}, dynamic polarization of electron spins~\cite{PhysRevLett.125.156801,PhysRevB.104.115405} \textit{e. t. c.}

In the past few years the interest in the investigation of spins in quantum dots has shifted from electrons and holes to the nuclei of the host lattice~\cite{waeber2019pulse,Gangloff62,chekhovich2020nuclear,jackson2021quantum}. This is related with their longer spin relaxation and dephasing times, which allows one to talk about nuclear spins as a platform for the storage and processing of the quantum information~\cite{PhysRevLett.90.206803}.

The simplest exactly solvable model for the description of the nuclear spin dynamics is the box model~\cite{Kozlov2007}. In the framework of this model, we have obtained transparent analytical expressions for the nuclear spin dynamics and have calculated analytically their spin noise spectra~\cite{PhysRevLett.126.216804}, which agree with the previous numeric calculations~\cite{PhysRevB.97.195311}.

In this work, we rederive the previous results by a new method, and also generalize them for the case of the anisotropic hyperfine interaction. An important result is the prediction of the nuclear spin dynamics for anisotropic hyperfine interaction in zero magnetic field. In particular, this prediction opens the way for the efficient detection of the current noise induced by the nuclear spin noise in organic semiconductors.

\section{Exact solution of box model}

The general Hamiltonian of the central spin model in an external magnetic field has the form
\begin{equation}
  \mathcal H=\sum_{k=1}^N\bm{S}\hat{\bm A}_k\bm I_k+\hbar\bm\Omega_B\bm S+\sum_{k=1}^N\hbar\bm\omega_B^{(k)}\bm I_k.
\end{equation}
Here $\bm S$ is the central spin, $k$ enumerates $N$ nuclear spins $\bm I_k$, $\hat{\bm A}_k$ are the corresponding tensors of the hyperfine interaction, $\bm\Omega_B$ and $\bm\omega_B^{(k)}$ are the Larmor precession frequencies of electron and nuclear spins in the external magnetic field, respectively. We consider, as usual, all nuclear and electron spins equal to $1/2$.

In this work, we consider the box model, where all the hyperfine interaction tensors and all the nuclear spin precession frequencies are equal: $\hat{\bm A}_k=\hat{\bm A}$ and $\bm\omega_B^{(k)}=\bm\omega_B$. In this case, the Hamiltonian can be written it terms of the total nuclear spin
\begin{equation}
  \bm I=\sum_{k=1}^N\bm I_k:
\end{equation}
\begin{equation}
  \label{eq:Ham}
  \mathcal H=\bm{S}\hat{\bm A}\bm I+\hbar\bm\Omega_B\bm S+\hbar\bm\omega_B\bm I.
\end{equation}
We consider the case, when the hyperfine interaction tensor has the nonzero components $A_{zz}\equiv A_\parallel$ and $A_{xx}=A_{yy}=A_\perp$, and the external magnetic field is applied along the $z$ axis, so $\bm\Omega_B$ and $\bm\omega_B$ are parallel to this axis. The spectrum of this Hamiltonian was found by Kozlov~\cite{Kozlov2007}, and below we reproduce his result.

The Hamiltonian of the box model conserves the absolute value of the total nuclear angular momentum, so the eigenstates can be parameterized by the quantity $I$, which takes the values from $(N~\rm{mod}~2)/2$ to $N/2$ with the step $1$. For the given value of $I$, the system has $2(2I+1)P$ eigenstates, where the multiplier $2$ describes the number of electron spin states, $2I+1$ is the number of the total nuclear spin projections, and
\begin{equation}
  P=C_{N}^{N/2+I}-C_{N}^{N/2+I-1}
\end{equation}
is the number of the realizations of the given nuclear spin $I$ expressed as the difference of the binomial coefficients~\cite{ll3_eng}.

In the system, the projection of the total angular momentum $F_z=S_z+I_z$ is conserved. It takes the values from $-I-1/2$ to $I+1/2$ with the step $1$. For each of the values except for the largest and the smallest, there are two eigenstates, which are the linear combinations of the functions $\ket{F_z-1/2,\uparrow}$ and $\ket{F_z+1/2,\downarrow}$, where $\ket{I_z,\uparrow/\downarrow}$ are the wave functions with the nuclear spin projection $I_z$ along the $z$ axis and the electron spin parallel/antiparallel to the $z$ axis. In the basis of this pair of states, the Hamiltonian can be written as
\begin{equation}
  \mathcal H(F_z)=\hbar\omega_B F_z-\frac{A_\parallel}{4}+\hbar\bm\Omega_e\bm S,
\end{equation}
where
\begin{subequations}
  \label{eq:Omega_e_exact}
  \begin{equation}
    \Omega_{e,x}=\frac{A_\perp}{\hbar}\sqrt{I(I+1)-F_z^2+1/4},
  \end{equation}
  \begin{equation}
    \Omega_{e,y}=0,
    \qquad
    \Omega_{e,z}=\Omega_B-\omega_B+\frac{A_\parallel}{\hbar}F_z.
  \end{equation}
\end{subequations}
The quantity $\bm\Omega_e$ depends on $F_z$, but hereafter we omit this argument for brevity. Eigenenergies have the form
\begin{equation}
  \label{eq:E}
  E_\pm(F_z)=\hbar\omega_B F_z-\frac{A_\parallel}{4}\pm\hbar\Omega_e/2,
\end{equation}
where $\Omega_e=|\bm\Omega_e|=\frac{1}{\hbar}\sqrt{A_\perp^2I(I+1)+\hbar^2(\Omega_B-\omega_B)^2+2A_\parallel F_z\hbar(\Omega_B-\omega_B)+(A_\parallel^2-A_\perp^2)F_z^2+A_{\perp}^2/4}$, and the corresponding wave functions are
\begin{equation}
  \Psi_\pm(F_z)=\mathcal A_\pm(F_z)\ket{F_z+1/2,\downarrow}+\mathcal B_\pm(F_z)\ket{F_z-1/2,\uparrow},
\end{equation}
with
\begin{equation}
  \mathcal A_+(F_z)=-\mathcal B_-(F_z)=\frac{\Omega_{e,x}}{\sqrt{2\Omega_e(\Omega_e+\Omega_{e,z})}},
  \qquad
  \mathcal B_+(F_z)=\mathcal A_-(F_z)=\sqrt{\frac{\Omega_e+\Omega_{e,z}}{2\Omega_e}}.
\end{equation}
Moreover, there are two extreme eigenstates $\ket{N/2,\uparrow}$ and $\ket{-N/2,\downarrow}$ with the energies $A_\parallel N/4\pm\hbar(\Omega_B+\omega_BN)/2$, respectively.

For the description of the nuclear spin dynamics, let us consider the linear combinations of the wave functions
\begin{equation}
  \Psi_\pm(t)=\alpha_\pm(t)\Psi_\pm(I_z+1/2)+\beta_\pm(t)\Psi_\pm(I_z-1/2)
\end{equation}
for the given value of $I_z$. Time evolution of this combination is described by
\begin{equation}
  \label{eq:schr}
  \alpha_\pm(t)=\alpha_\pm(0)\exp[-\i E_\pm(I_z+1/2)t/\hbar],
  \qquad
  \beta_\pm(t)=\beta_\pm(0)\exp[-\i E_\pm(I_z-1/2)t/\hbar].
\end{equation}
The average transverse components of the nuclear spin are given by the expression $\braket{I_{x,y}(t)}=\braket{\Psi_\pm(t)|I_{x,y}|\Psi_\pm(t)}$. Explicitly, they are equal to
\begin{multline}
  \braket{I_{x}(t)+\i I_y(t)}=\alpha_\pm^*(t)\beta_\pm(t)\left[\mathcal A(I_z+1/2)\mathcal A(I_z-1/2)\sqrt{(I-I_z)(I+I_z+1)}
    \right.\\\left.
      +\mathcal B(I_z+1/2)\mathcal B(I_z-1/2)\sqrt{(I-I_z+1)(I+I_z)}\right].
\end{multline}
From Eq.~\eqref{eq:schr} we obtain
\begin{equation}
  \label{eq:rotation}
  \braket{I_{x}(t)+\i I_y(t)}=\braket{I_{x}(0)+\i I_y(0)}\e^{\i\omega_\pm t},
\end{equation}
where
\begin{equation}
  \label{eq:omega_pm_exact}
  \omega_\pm=\frac{E_\pm(I_z+1/2)-E_\pm(I_z-1/2)}{\hbar}.
\end{equation}
Thus, the average nuclear spin in the $(xy)$ plane for the linear combinations of the wave functions under consideration precesses about the $z$ axis with the frequency $\omega_\pm$.

\section{Large box limit}

In the limit of the large nuclear spin, $I\gg 1$, from Eq.~\eqref{eq:omega_pm_exact} we obtain
\begin{equation}
  \label{eq:omega_pm}
  \omega_\pm=\omega_B\pm\frac{A_\parallel\hbar\Omega_B+(A_\parallel^2-A_\perp^2)I_z}{2\hbar^2\Omega_e}
\end{equation}
where we neglect $\omega_B$ in comparison with $\Omega_B$ because of the large difference between electron and nuclear $g$-factors. Moreover, in this limit, from Eq.~\eqref{eq:Omega_e_exact} we obtain
\begin{equation}
  \label{eq:Omega_e}
  \bm\Omega_e=\bm\Omega_B+\frac{\hat{\bm A}}{\hbar}\bm I.
\end{equation}

These expressions can be also obtained from the following qualitative considerations. Let us consider $\bm I$ and $\bm S$ as classical three-dimensional vectors. Then, as one can see from the Hamiltonian~\eqref{eq:Ham}, $\bm\Omega_e$ from Eq.~\eqref{eq:Omega_e} is indeed the electron spin precession frequency for the given $\bm I$. Further, as one can see from Eq.~\eqref{eq:E}, the eigenstates with the subscripts $\pm$ correspond to the electron spin parallel and antiparallel to $\bm\Omega_e$. In these states,
\begin{equation}
  \label{eq:S_proj}
  \bm S=\pm\frac{\bm\Omega_e}{2\Omega_e}.
\end{equation}

In the same time, the nuclear spin precession frequency, as it also follows from the Hamiltonian~\eqref{eq:Ham}, equals to
\begin{equation}
  \bm\omega_n=\bm\omega_B+\frac{\hat{\bm A}}{\hbar}\bm S.
\end{equation}
For the states with the subscripts $\pm$, from Eq.~\eqref{eq:S_proj} we obtain
\begin{equation}
  \bm\omega_n=\bm\omega_B\pm\frac{\hat{\bm A}\bm\Omega_e}{2\hbar\Omega_e}
\end{equation}
Now, we substitute here Eq.~\eqref{eq:Omega_e} and use the fact that one can subtract from the precession frequency of $\bm I$ a vector $A_\perp^2\bm I/(2\hbar^2\Omega_e)$, which is parallel to $\bm I$. As a result, we obtain
\begin{equation}
  \frac{\d\bm I}{\d t}=\omega_\pm\bm e_z\times\bm I,
\end{equation}
where $\omega_\pm$ is given by the Eq.~\eqref{eq:omega_pm}. In this expression, the precession frequency $\omega_\pm$ does not depend on time.

\begin{figure}[t]
  \centering
  \includegraphics[width=0.5\linewidth]{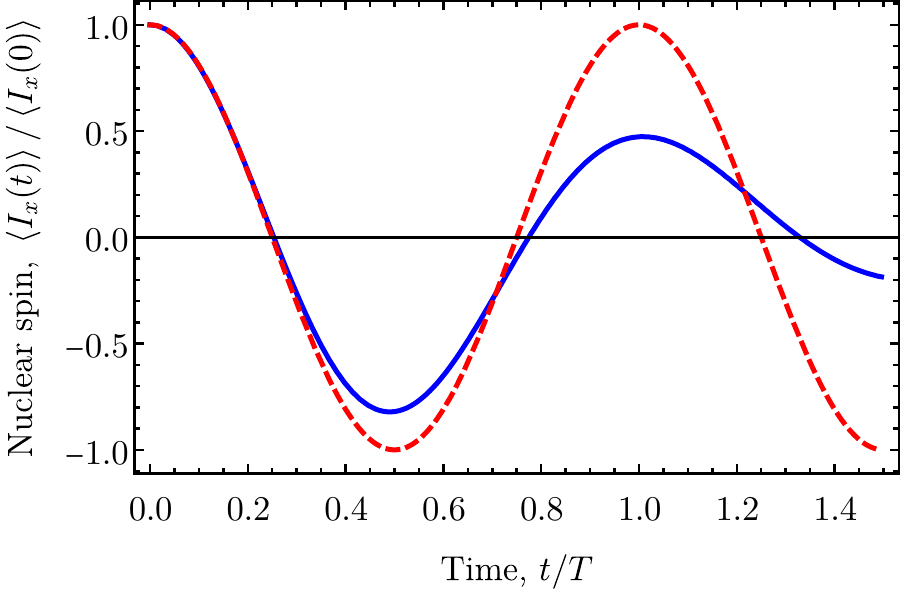}
  \caption{The numeric calculation of $\braket{I_x(t)}$ for the cases of $A_\parallel=0$ (blue solid curve) and $A_\perp=0$ (red dashed curve) and for the parameters given in the main text. The period of oscillations is determined from Eq.~\eqref{eq:omega_pm} as $T=2\pi/|\omega_\pm|$.}
  \label{fig:osc}
\end{figure}

Figure~\ref{fig:osc} shows the exact calculation of the oscillations of $\braket{I_x(t)}$ in zero magnetic field, $\Omega_B=\omega_B=0$, for the cases of anisotropic hyperfine interaction $A_\perp=0$ and $A_\parallel=0$. As an initial condition we chose the orientation of the total nuclear spin $I=50$ in the $(xz)$ plane with the angle $\pi/4$ to the axes and unpolarized electron spin. One can see that the nuclear spin indeed precesses with the frequency~\eqref{eq:omega_pm}. For the limit $A_\parallel=0$ one can see also the decay of the oscillations, which disappears in the limit of large $I$.

\section{Nuclear spin noise}

One of the most powerful methods to investigate the nuclear spin dynamics is the spin noise spectroscopy~\cite{Smirnov-review,PhysRevB.101.235416,NuclearNoise,OpticalField,PolarizedNuclei}. The nuclear spin noise spectrum is defined as
\begin{equation}
  \label{eq:spec_def}
  (I_\alpha^2)_\omega=\int\limits_{-\infty}^\infty\braket{I_\alpha(0)I_\alpha(\tau)}\e^{\i\omega\tau}\d\tau,
\end{equation}
where $\alpha=x,y,z$ and the angular brackets denote the quantum statistical averaging. We consider the experimentally relevant case of high temperatures, when the nuclear polarization is negligible, so the distribution function of the total nuclear spin has the form~\cite{Smirnov-review}
\begin{equation}
  \label{eq:F}
  \mathcal F(\bm I)=\left(\frac{2}{\pi N}\right)^{3/2}\exp\left(-\frac{2I^2}{N}\right).
\end{equation}

The dynamics of the nuclear spin $z$ component is absent in our model, while for the two other components from the axial symmetry one has
\begin{equation}
  (I_x^2)_\omega=(I_y^2)_\omega.
\end{equation}
So in the following we consider the spectrum $(I_x^2)_\omega$. Moreover, we neglect the Zeeman splitting of the nuclear spin sublevels $\omega_B$, because it leads to the splitting of the spectrum into the two equivalent components only~\cite{PhysRevLett.126.216804}.

In this limit, it follows from the equation of motion~\eqref{eq:rotation} that
\begin{equation}
  \braket{I_x(0)I_x(\tau)}=\braket{I_x^2(0)\cos(\omega_n\tau)},
\end{equation}
where
\begin{equation}
  \label{eq:omega_n}
  \omega_n=\left|\frac{A_\parallel\hbar\Omega_B+(A_\parallel^2-A_\perp^2)I_z}{2\hbar\sqrt{(\hbar\Omega_B+A_\parallel I_z)^2+A_\perp^2(I_x^2+I_y^2)}}\right|
\end{equation}
is derived from Eq.~\eqref{eq:omega_pm}. Further, from the definition~\eqref{eq:spec_def} we obtain the general expression for the spectrum at positive frequencies
\begin{equation}
  \label{eq:spec_general}
  (I_x^2)_\omega=\pi\braket{I_x^2\delta(\omega-\omega_n)},
\end{equation}
where the averaging should be performed with the distribution function~\eqref{eq:F}.

We could not derive a general analytical expression for the spin noise spectrum. Therefore in what follows we consider a number of particular cases.

The most interesting case is the limit of zero magnetic field, $\Omega_B=0$. In this limit, the nuclear spin precession frequency~\eqref{eq:omega_n} does not depend on the absolute value of the nuclear spin:
\begin{equation}
  \omega_n=\frac{\left|(A_\parallel^2-A_\perp^2)I_z\right|}{2\hbar\sqrt{A_\parallel^2I_z^2+A_\perp^2(I_x^2+I_y^2)}}.
\end{equation}
The averaging over the directions of $\bm I$ gives
\begin{equation}
  \label{eq:B0}
  (I_x^2)_\omega=\frac{3}{4}\pi N\hbar A_\perp\frac{A_\perp^4+A_\parallel^4-2A_\parallel^2\left[A_\perp^2+2\left(\hbar\omega\right)^2\right]}{\sqrt{\left|A_\parallel^2-A_\perp^2\right|}\left|A_\parallel^2-A_\perp^2-(2\hbar\omega)^2\right|^{5/2}}.
\end{equation}
Here, the spectrum is limited by the frequency $|A_\parallel^2-A_\perp^2|/(2\hbar A_\parallel)$.

If the longitudinal component of the hyperfine interaction tensor is zero, $A_\parallel=0$, then this expression simplifies to
\begin{equation}
  \label{eq:B0_perp}
  (I_x^2)_\omega=\frac{3\pi\hbar N}{4A_\perp\left[(2\hbar\omega/A_\perp)^2+1\right]^{5/2}}.
\end{equation}
Here the spectrum extends to all frequencies.

\begin{figure}
  \includegraphics[width=0.49\linewidth]{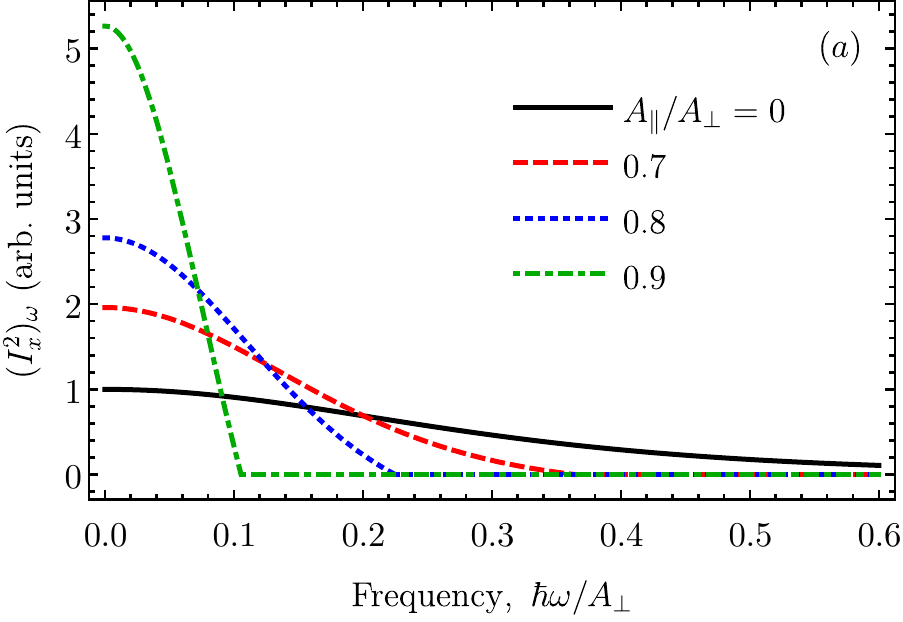}
  \includegraphics[width=0.49\linewidth]{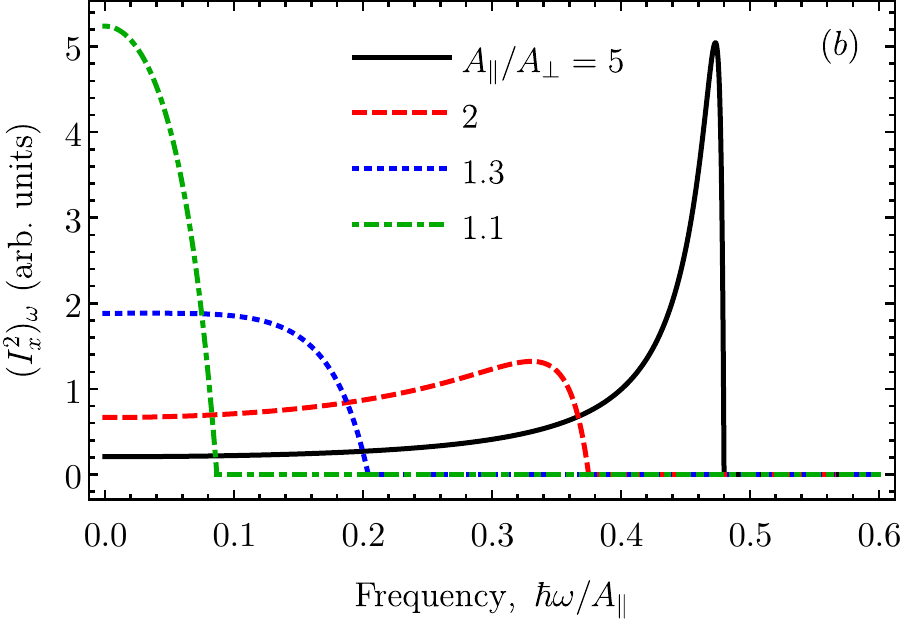}
  \caption{The spin noise spectra in zero magnetic field, $\Omega_B=0$, for the different anisotropies of the hyperfine interaction indicated in the legends, calculated after Eq.~\eqref{eq:B0}.}
  \label{fig:A_dep}
\end{figure}

In Fig.~\ref{fig:A_dep} we show the spin noise spectra for the different degrees of the anisotropy. For convenience, the frequency is normalized by $A_\perp/\hbar$ in the case of $A_\perp>A_\parallel$ [panel (a)] and by $A_\parallel/\hbar$ in the case of $A_\perp<A_\parallel$ [panel (b)]. In the limit of the hyperfine interaction in $(xy)$ plane, $A_\parallel=0$, the spectrum is centered at zero frequency and has the width of the order of $A_\perp/\hbar$ in agreement with Eq.~\eqref{eq:B0_perp}. With decrease of the anisotropy, the spectrum narrows remaining centered at zero frequency until $A_\parallel=A_\perp$, when the spectrum becomes singular.

Further, with increase of the ratio $A_\parallel/A_\perp$ the maximum in the spectrum appears at a finite frequency, which shifts towards $\omega=A_\parallel/(2\hbar)$ in the limit of $A_\parallel\gg A_\perp$. In the same time, the width of the spectrum first increases and then decrease and eventually vanishes. This is related to the fact that the nuclear spin precession frequency equals to $A_\parallel/(2\hbar)$, as it follows from Eq.~\eqref{eq:omega_n}, and it does not depend on the direction of the nuclear spin $\bm I$.

Let us turn to the analysis of the role of the magnetic field.

First, if the hyperfine interaction is predominantly along the $z$ axis, $A_\parallel\gg A_\perp$, then the spin precession frequency remains equal to $A_\parallel/(2\hbar)$ and the spin noise spectrum does not get modified.

Second, in the case of the isotropic hyperfine interaction, $A_\parallel=A_\perp$, we have shown previously that the spectrum shifts from the zero frequency to $A/(2\hbar)$ with increase of the magnetic field~\cite{PhysRevLett.126.216804}. In the same time, its width changes nonmonotonously and tends to zero in the limit of $\Omega_B\to\infty$.

Finally, in the limit of the hyperfine interaction predominantly in the $(xy)$ plane, $A_\parallel=0$, one can perform averaging in the general Eq.~\eqref{eq:spec_general} analytically, first over $I_z$ with the help of the $\delta$-function, and then over the components $I_x$ and $I_y$. The answer for the spin noise spectrum in this limit has the form
\begin{equation}
  \label{eq:spec_perp}
  (I_x^2)_\omega=\frac{\sqrt{\pi}\hbar N}{4A_\perp(1+\nu^2)^{5/2}}\left[6b\sqrt{1+\nu^2}\e^{-\nu^2b^2}+\sqrt{\pi}\left(3-2b^2-2b^2\nu^2\right){\rm Erfc}(b\sqrt{1+\nu^2})\e^{b^2}\right],
\end{equation}
where for the brevity we introduce the dimensionless frequency $\nu=2\hbar\omega/A_\perp$ and the dimensionless magnetic field $b=(\hbar\Omega_B/A_\perp)\sqrt{2/N}$, with ${\rm Erfc}(x)=1-{\rm Erf}(x)$ being the complementary error function. In the limit of zero magnetic field, $\Omega_B\to0$, this expression transforms into previously derived answer~\eqref{eq:B0_perp}.

\begin{figure}
  \includegraphics[width=0.5\linewidth]{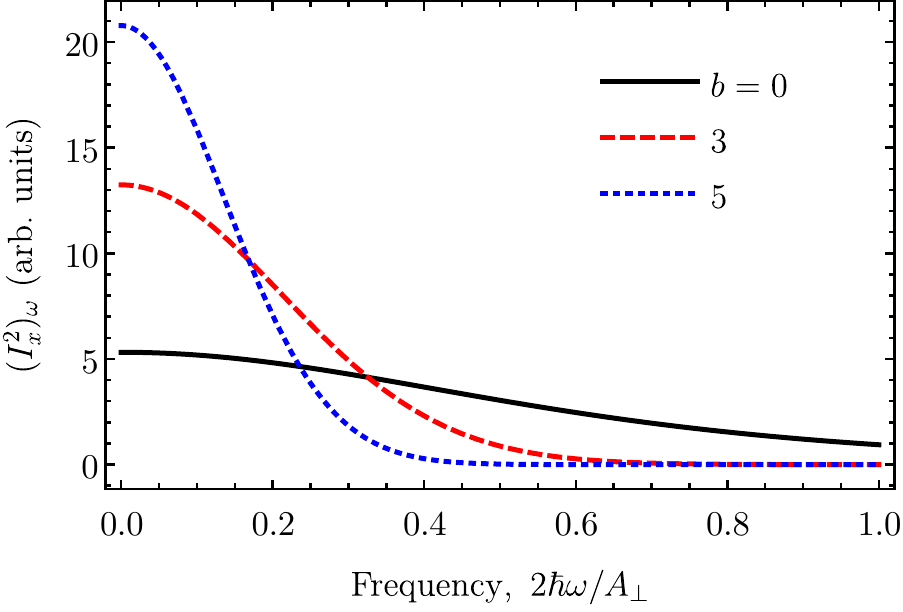}
  \caption{Nuclear spin noise spectra for the hyperfine interaction in the $(xy)$ plane, $A_\parallel=0$, for the different magnetic fields indicated in the legend, calculated after Eq.~\eqref{eq:spec_perp}.}
  \label{fig:B_dep}
\end{figure}

The dependence of the spectra on the magnetic field in this limit is illustrated in Fig.~\ref{fig:B_dep}. With increase of the magnetic field, the spectrum narrows, but remains centered at zero frequency.

\begin{figure}[t]
  \centering
  \includegraphics[width=0.49\linewidth]{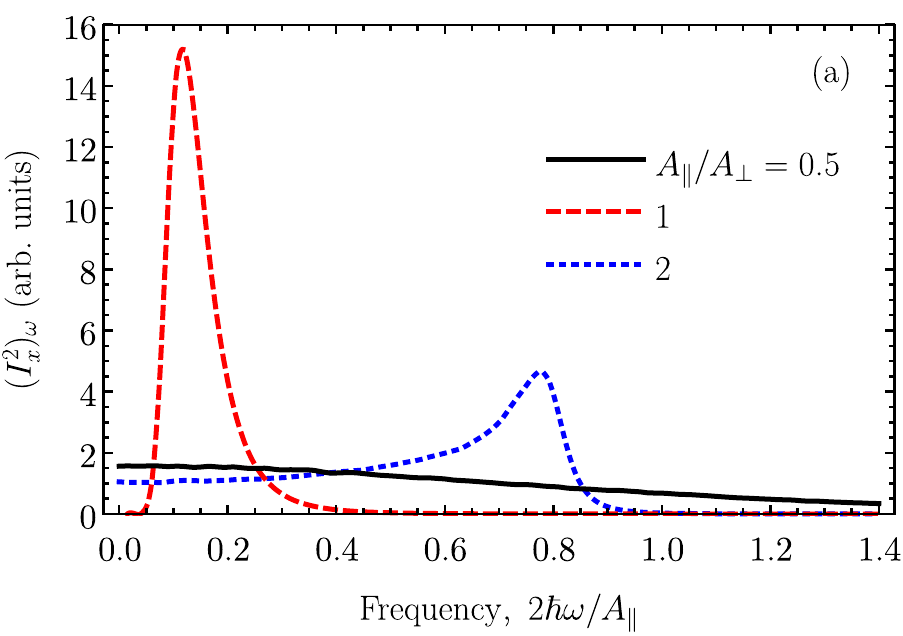}
  \includegraphics[width=0.49\linewidth]{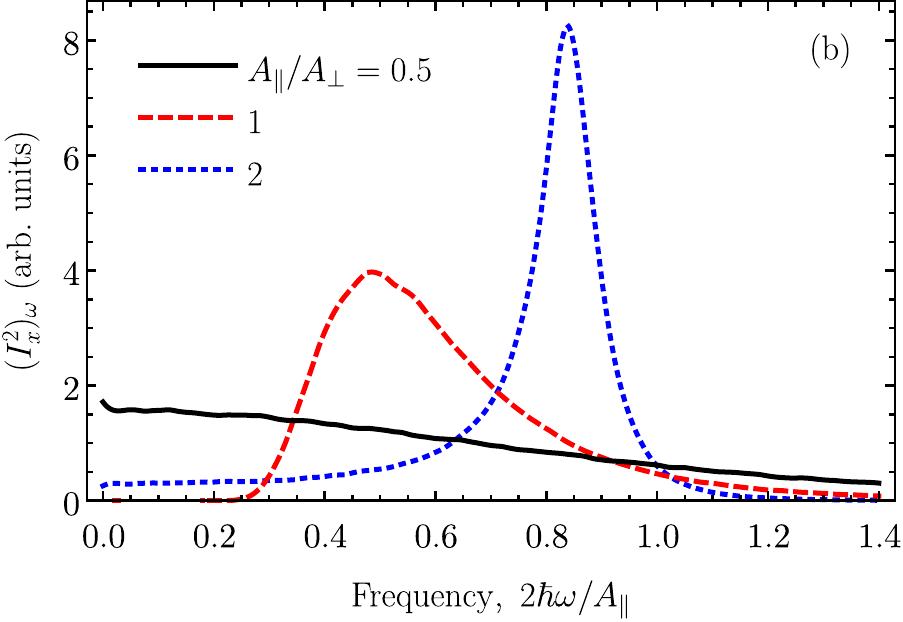}
  \includegraphics[width=0.49\linewidth]{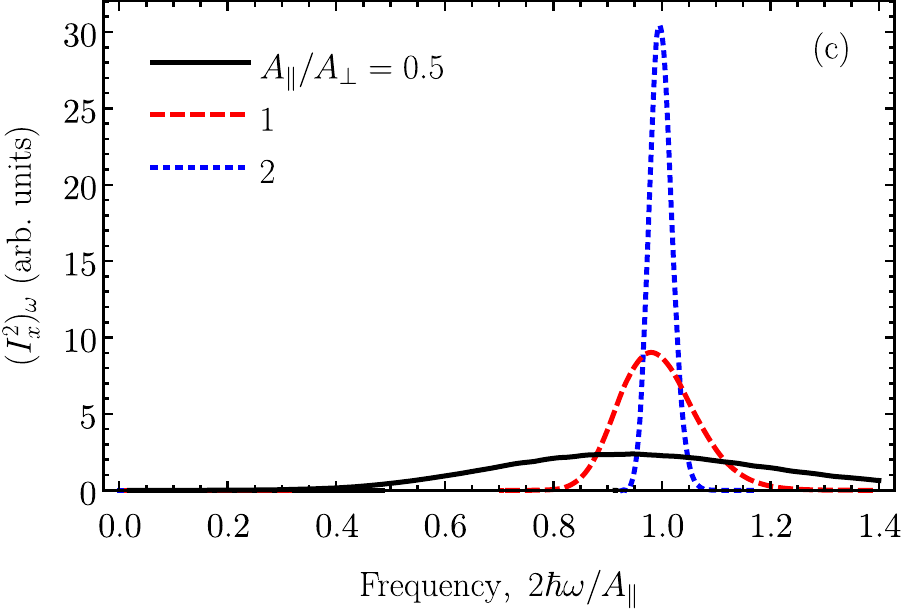}
  \caption{The spin noise spectra, calculated numerically after Eq.~\eqref{eq:spec_general}, for the different magnetic fields $(\hbar\Omega_B/A_\parallel)\sqrt{2/N} = 0.2$ (a), $1$ (b) and $10$ (c) for the anisotropies of the hyperfine interaction given in the legends.}
  \label{fig:noise}
\end{figure}

Generally, the spin noise spectra can be calculated numerically after Eq.~\eqref{eq:spec_general}. In Fig.~\ref{fig:noise} we show the results of the numeric calculations with the averaging over  $10^7$ random initial nuclear spin states. The three values of the magnetic field are shown: $\hbar\Omega_B = 0.2A_{\parallel}\sqrt{N/2}$, when the Zeeman energy is smaller than the hyperfine interaction energy; $\hbar\Omega_B = A_{\parallel}\sqrt{N/2}$ when they are comparable; and $\hbar\Omega_B = 10A_{\parallel}\sqrt{N/2}$, when the former dominates the latter. For each value of the field the three degrees of the anisotropy are considered: $A_\parallel/A_\perp=2$, $1$ and $1/2$.

In particular, Fig.~\ref{fig:noise}(a) confirms the nuclear spin precession in small magnetic field. Here in agreement with the limiting cases described above, the peak in the spectrum is narrow and is located at a low frequency, while for the cases of $A_\parallel/A_\perp=2$ and $1/2$ the peak is broad and is located at finite and zero frequencies, respectively.

It follows from Fig.~\ref{fig:noise}(с) that in the large magnetic fields the spectrum, in agreement with the previous analysis, is centered at the frequency $A_\parallel/(2\hbar)$ and its width is determined by $A_\perp$. Panel (b) in Fig.~\ref{fig:noise} shows the transition between the two limiting cases.

\section{Discussion and conclusion}

The description of the nuclear spin dynamics from the exact Hamiltonian diagonalization was performed in this work for the states, which are coherent superpositions of two eigenfunctions. In reality, the nuclear spin state with the given average value of $\bm I$ typically is a coherent collective nuclear spin state, i.e. a superposition of  $\sim\sqrt{I}$ eigenfunctions. However, for the large $I$, $\sqrt{I}$ is relatively small, so the results derived above are valid.

Moreover, in the framework of this model one can consider also coherent superpositions of the states with the different orientations of the electron spin. However, since the nuclear spin dynamics is slow, the electron spin coherence in realistic systems at the corresponding time scales, would not be preserved, probably. Nevertheless, this does not affect the applicability of our theory to the description of the nuclear spin dynamics, nuclear spin noise, and also generation of the entangled and squeezed nuclear spin states~\cite{PhysRevLett.126.216804} at the time scales up to the longitudinal electron spin relaxation time. In the quantum dots of the A$_3$B$_5$ type, these times are typically of the order of a few microseconds.

A significant disadvantage of the considered model is the approximation of the homogeneous hyperfine interaction, which is typically not fulfilled. Nevertheless, the previous investigations point out that the box model is qualitatively correct~\cite{PhysRevLett.91.017402,PhysRevLett.104.143601,PhysRevB.97.195311,PhysRevB.99.035439,PhysRevB.102.085413}, in particular, for the subensemble of the nuclei in the vicinity of the center of the electron wave function, the constants of the hyperfine interaction are close to each other. Also, an interesting generalization of the model would be the consideration of the different orientations of the principal axis of the hyperfine interaction and external magnetic field.

One of the interesting applications of the developed theory is the description of the electric properties of organic semiconductors. It turns out that in many of them, the hyperfine interaction between the electron spins and the nuclei of the molecules significantly affects the conductivity even at the room temperature~\cite{https://doi.org/10.1002/pssb.201046383}. Previously, we have shown that the nuclear spin noise in these materials leads to the current noise~\cite{noise_OMAR}. Its measurement would give an important information about the details of the hyperfine interaction in organic semiconductors.

Such a measurement, however, is spoiled by other sources of the low frequency noise such as, for example, the broad distribution of the probabilities of the electron hops between the molecules~\cite{PhysRevB.95.115204}. We believe that it is easier to separate the contribution related to the nuclear spin fluctuations in the case when it represents a peak at the finite frequency, which is larger than its width. Previously, we have shown that this situation is realized for the isotropic hyperfine interaction in the strong enough magnetic field, $\hbar\Omega_B\gtrsim\sqrt{N}A$. In this case, however, the influence of the nuclear spins on the conductivity is suppressed and the amplitude of the current fluctuations, induced by the nuclear spin noise, is small.

But the result of this work stands that the spin noise spectrum has a peak at a finite frequency even in zero magnetic field, if the hyperfine interaction is more efficient along the $z$ axis, $A_\parallel>A_\perp$. This regime, as we think, is the most perspective for the experimental observation of the current noise, induced by the nuclear spin noise. In the same time, this very regime is expected because of the mostly $p$-type of the electronic states in the organic molecules. For these states, the contribution of the dipole-dipole hyperfine interaction turns out to be comparable to the contribution from the isotropic contact interaction~\cite{PhysRevB.87.205446}. The specific form of the interaction tensor along with the possibility to apply the box model are defined by the configuration of molecules and should be studied separately for each organic semiconductor.

In conclusion, in this work from the exact diagonalization of the Hamiltonian of the box model we obtained the analytic expressions for the nuclear spin dynamics for many nuclear spins accounting for the anisotropy of the hyperfine interaction. These expressions allowed us to calculate the nuclear spin noise spectra and to obtain for them the analytical expressions in the number of limiting cases. The key result is the prediction of the nuclear spin precession in zero magnetic field, which can be used for the detection of the nuclear induced current noise in organic semiconductors.

\section*{Acknowledgements}

We thank \href{https://orcid.org/0000-0003-3607-7539}{L. Lanco} for useful discussions, the organizing committee of the \href{http://www.ioffe.ru/coherent/sstm2021/Sovesanie.html}{Workshop on Solid State Theory} for the motivation to write the manuscript, and also the Foundation for the Advancement of Theoretical Physics and Mathematics ``BASIS,'' RF President Grant No. MK-5158.2021.1.2 and RFBR grant No. 19-02-00184 for partial financial support. The analytic calculations of the spin noise spectra by D.S.S. were supported by the Russian Science Foundation Grant No. 21-72-10035.


%

\end{document}